\documentstyle[12pt,a4,epsf]{article}
\begin{document}
\begin{flushright}
PSI-PR-97-17 
\\ July 14, 1997 
\end{flushright} 

\begin{center}
{\LARGE OZI Rule Violation in $p\bar{p}$ Annihilation into $\phi\pi\pi$ \\
        by Two Step Processes}
\\[5mm]
{\sc V.E.~Markushin and M.P.~Locher}
\\[5mm]
{\it Paul Scherrer Institute, 5232 Villigen PSI, Switzerland}
\end{center}

\vspace*{8mm}
\begin{abstract}

The $\phi\pi^+\pi^-$ production in $p\bar{p}$ annihilation at rest is
strongly enhanced by a two step mechanism with intermediate
$K\bar{K}\pi\pi$ states. The relative yield of the $\phi$ production due
to the resonant final state interaction decreases with increasing total
energy of the $p\bar{p}$ system. 

\end{abstract}

\vspace*{10mm}

\section{Introduction}
\label{Intr}

  The production of $\phi$ mesons in low energy $p\bar{p}$ annihilation  (see
\cite{AST91,OBEL1,OBEL2,CBC95,JETSET95a,OBEL96} and references therein) is
strongly enhanced in some channels for which the $\phi$ production is
expected to be suppressed on the tree level according to the
Okubo-Zweig-Iizuka (OZI) rule \cite{Ok63,Zw64,Ii66}. These conspicuously OZI
rule violating reactions are of considerable theoretical interest since they
can be used for studying reaction mechanisms and the nucleon structure,
in particular the problem of an intrinsic $s\bar{s}$ component 
\cite{EGK,EKKS,GYF}.  
  Some well-known OZI-rule breaking mechanisms are two-step
processes with ordinary hadrons in the intermediate state \cite{Li84},
therefore their role in the $p\bar{p}$ annihilation must be 
investigated.  

  The two step mechanisms in nucleon-antinucleon annihilation have been
studied for various final states containing $\phi$ mesons
\cite{LZL,LLZ,LL,BL94,BL94a,MHS,BL95,BL95a,GLMR,AK96}. It has been 
found that in the $\phi\pi$ and $\phi\phi$ channels, which show the most
dramatic violation of the OZI rule, two-step mechanisms play an important
role and result in a cross section comparable with the experimental data.  
  However, for a number of reactions, including
$p\bar{p}\to\phi\rho,\;\phi\pi\pi,\;\phi\omega$, no significant
contributions from two-step mechanisms have been found. 
  The goal of this paper is to reanalyze the role of two-step mechanisms for
the $p\bar{p}$ annihilation into the $\phi\pi\pi$ channel. To this aim we
go beyond the approximations used in previous studies. 
The resulting cross sections turn out to have the right order of magnitude.
The plan of the paper is as follows. Section~\ref{EXP} gives a
brief summary of the experimental data and the previous calculations.
  A simple model illustrating the two-step mechanism 
with a resonant final state interaction is considered in Sec.~\ref{RRM}.   
The reaction $p\bar{p}\to\phi\pi\pi$ is discussed in Sec.~\ref{RRMphipipi}
where the two-step mechanism with the $K\bar{K}\pi\pi$ intermediate state is
studied in detail. The summary of the results is given in Sec.~\ref{CONCL}. 
For earlier reviews concerning the two-step mechanisms we refer to
\cite{Lo95,Sa95,Zou96,Ma96,Wi96}. 

\section{The OZI Rule Violation in
         $p\bar{p}\to\phi\pi^+\pi^-,\phi\rho,\phi\omega$}
\label{EXP}

Table~\ref{OZIphirho} gives a summary of the experimental data on the OZI rule
violation in the $\phi\pi^+\pi^-$ and $\phi\rho$ channels at different
energies. The tree level expectations based on the deviation of
the $\omega-\phi$ mixing angle from the ideal one are
$(\Theta-\Theta_i)^2\sim 4\cdot 10^{-3}$ \cite{PDT}.  The OZI rule
violation in these channels is therefore rather moderate\footnote{%
This magnitude of the OZI rule breaking is often termed nondramatic,
contrary to the cases where the $\phi$ production exceeds the
estimate from $\phi-\omega$ mixing by more than one order of magnitude.}.  
The experimental branching ratios for the reactions
$p\bar{p}\to \pi\pi\phi,\ \rho\phi,\ \pi\pi\omega, \rho\omega,\
\pi\pi K\bar{K}$ at rest are given in Tab.~\ref{TabpipiKK}.

\begin{table}[hbt]
\caption{\label{OZIphirho} The relative probabilities for
the $\phi\rho,\omega\rho$ and $\phi\pi^+\pi^-,\omega\pi^+\pi^-$
channels in $p\bar{p}$ annihilation {\it vs.} antiproton momentum. 
Phase space corrected ratios of yields are also tabulated.} 
\vspace*{2mm}
\begin{center}
\begin{tabular}{|l|c|c|c|c|l|}
\hline\hline
$p_{\bar{p}}$(GeV/c) &
\multicolumn{2}{|c|}%
{$\frac{BR(\phi\rho)}{BR(\omega\rho)}\cdot 10^3$} &
\multicolumn{2}{|c|}%
{$\frac{BR(\phi\pi^+\pi^-)}{BR(\omega\pi^+\pi^-)}\cdot 10^3$} &
Ref. \\
\cline{2-5} 
  &  measured & PS corrected & measured & PS corrected &   \\
\hline
 0 (gas)    & $6.3 \pm 1.6$ &       &         &        &
\cite{AST91,AST93} \\
 0 (gas/LX) & $7.5 \pm 2.4$ &       &         &        &
\cite{AST91,AST93} \\
 0 (liq.)   &       &        & $7.0 \pm 1.4$  &  $15 \pm 3$     &
\cite{Biz69} \\
 0 (liq.)   &       &        & $4.9 \pm 0.8$  & $10.3 \pm 1.6$  &
\cite{OBEL96} \\
 0 (3 atm)  &       &        & $5.9 \pm 0.9$  & $12.5 \pm 2.0$  &
\cite{OBEL96} \\
 0.76       & $9 \pm 5$ & $13 \pm 4$  &  $10.0\pm 2.4$ & $19 \pm 5$  &
\cite{Co78} \\
 1.2        &       &        & $11 {+3 \atop -4}$ & $19 {+5 \atop -7}$ &
\cite{Do76} \\
 2.3        & $22 \pm 13$ & $25 \pm 15$  &  $21 \pm 5$ & $30 \pm 7$  &
\cite{Ch77} \\
 3.6        &       &        & $9 {+4\atop -7}$ & $12 {+5\atop -9}$  &
\cite{Do76} \\  
\hline\hline
\end{tabular}
\end{center}
\end{table}

\begin{table}[hbt]
\caption{\label{TabpipiKK} The experimental branching ratios for $p\bar{p}$
annihilation at rest into
$\pi\pi K\bar{K}$, $\rho\phi$, and $\pi\pi\phi$.}
\begin{center}
\begin{tabular}{|c|c|c|l|}
\hline\hline
 Reaction &    BR   &   Condition  &  Ref. \\
\hline\hline
$p\bar{p}\to\pi^+\pi^- \phi$
                        & $4.6(9)\cdot 10^{-4}$  & liq. & \cite{Biz69} \\ 
                        & $5.4(10)\cdot 10^{-4}$ & gas  & \cite{AST91} \\ 
                        & $7.7(17)\cdot 10^{-4}$ & gas LX  & \cite{AST91} \\ 
                        & $4.7(11)\cdot 10^{-4}$ & $S$  & \cite{AST91} \\ 
                        & $6.6(15)\cdot 10^{-4}$ & $P$  & \cite{AST91} \\ 
                        & $3.5(4)\cdot 10^{-4}$  & liq. & \cite{OBEL96} \\ 
                        & $3.7(5)\cdot 10^{-4}$  & gas  & \cite{OBEL96} \\ 
\hline
$p\bar{p}\to\pi^+\pi^- \phi_{\to K_S K_L}$
                        & $1.8(3)\cdot 10^{-4}$  & liq. & \cite{ArREV} \\ 
\hline
$p\bar{p}\to\rho\phi$
                        & $3.4(8)\cdot 10^{-4}$   & gas. & \cite{AST91} \\
                        & $4.4(12)\cdot 10^{-4}$  & gas./LX & \cite{AST91} \\
                        & $3.4(10)\cdot 10^{-4}$  & $^1S_0$ & \cite{AST91} \\
                        & $3.7(9)\cdot 10^{-4}$   & $^3P_J$ & \cite{AST91} \\
\hline\hline
$p\bar{p}\to\pi^+\pi^- \omega$
                        & $6.6(6)\cdot 10^{-2}$   & liq. & \cite{Biz69} \\ 
\hline
$p\bar{p}\to\rho\omega$
                        & $5.4(6)\cdot 10^{-2}$  & gas. & \cite{AST93} \\
                        & $3.0(7)\cdot 10^{-2}$  &  S   & \cite{AST93} \\
                        & $6.4(11)\cdot 10^{-2}$ &  P   & \cite{AST93} \\
                        & $2.3(2)\cdot 10^{-2}$  & liq. & \cite{Biz69} \\
\hline\hline
$p\bar{p}\to\pi^+\pi^- K_S K_L$
                        & $2.41(36)\cdot 10^{-3}$ & liq. & \cite{Ba66} \\ 
                        & $2.26(45)\cdot 10^{-3}$ & liq. & \cite{ArREV} \\ 
\hline\hline
\end{tabular}
\end{center}
\end{table}


Until now the importance of the two-step mechanisms in the reactions
$p\bar{p}\to\phi\pi\pi$, $p\bar{p}\to\phi\rho$, $p\bar{p}\to\phi\omega$
remained unclear. The main attention was focused on two-particle
intermediate states (two meson doorway approximation). 
The contribution of the $K^*\bar{K}+K\bar{K}^*$
intermediate state was considered in the unitarity approximation in
\cite{BL94}; the calculated branching ratios for the $\phi\rho$ and
$\phi\omega$ channels were found to be about two orders of magnitude smaller
than observed.  The possibility of a large contribution of the $\rho\omega$
intermediate state for the $\phi\pi\pi$ channel was pointed out in
\cite{LL}, however the estimate was done in unitarity approximation and 
neglecting spin.
  Since the unitarity approximation is likely to be suppressed by threshold
factors, the off-mass-shell contributions can be large, but they are also 
known to be model dependent. Intermediate states with more than two
particles should also be taken into account.

  Of special interest are the intermediate states containing $K\bar{K}$,  
because the $K\bar{K}$ production is not OZI suppressed and the final state
interaction (FSI) effects are strong if the $K\bar{K}$ system is produced in
the region of the $\phi$ resonance.  The two-step mechanism
$p\bar{p}\to\pi\pi K\bar{K}\to\pi\pi\phi$ was estimated as well in \cite{LL}
using the unitariry approximation and neglecting the spin structure of the
amplitude, the resulting branching ratio being about one order  of magnitude
smaller than the experimental value.  In this paper we present a more
detailed calculation of this mechanism which includes spin effects and
off-shell contributions.  It shows that  $K\bar{K}$ rescattering in the
$K\bar{K}\pi\pi$ system  leads to a significant enhancement of the
$\phi\pi\pi$ production in $p\bar{p}$ annihilation at low energies which has
the right order of magnitude.  Note that this straight $K\bar{K}-\phi$
rescattering mechanism cannot contribute to two body final states in the two
meson doorway approximation.

\section{Resonant Rescattering Mechanism}
\label{RRM}

\begin{figure}
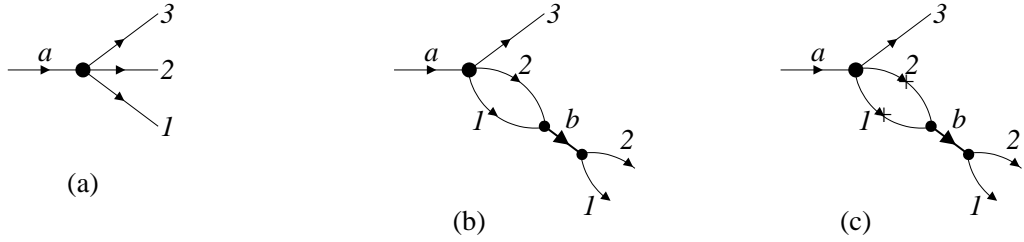

\begin{center}
\mbox{
\mbox{\epsfxsize=50mm\epsffile{Va123.epsf}}
\mbox{\epsfxsize=50mm\epsffile{Va123b12.epsf}}
\mbox{\epsfxsize=50mm\epsffile{Va123b12ua.epsf}}
   }
\end{center}
\caption{\label{DecayFSI}
(a) The amplitude $T_{a\to 123}^{0}$ for the decay $a\to 123$ in plane
    wave approximation, 
(b) The decay amplitude $T_{a\to 123}^{res}$ with resonant final
    state interaction. 
(c) The unitarity approximation $T_{a\to 123}^{UA}$ for the resonant FSI
    amplitude.} 
\end{figure}

\begin{figure}
\begin{center}
\mbox{\epsfxsize=60mm\epsffile{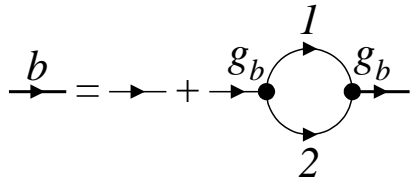}}
\end{center}
\caption{\label{MO}
The equation for the propagator of resonance $b$
coupled to the two-particle channel $(1+2)$.} 
\end{figure}

  To illustrate the basic features of the resonant rescattering mechanism
we consider a three-particle decay $a\to 123$ where particles 1 and 3
interact via resonance $b$. 
The total decay amplitude corresponds to the sum of the two diagrams shown
in Figs.\ref{DecayFSI}a,b.
As shown in Fig.\ref{MO}, all rescattering terms $1+2\leftrightarrow b$ are
taken into account by including the mass operator $\Pi_b(p_b^2)$ in the
resonance propagator. Here and below $p_n$ denotes the
four-momentum of particle $n$, $p_b^2=(p_1+p_2)^2=s_{12}$. 
The imaginary part of the mass operator $\Pi_b(s_{12})$ is 
determined by the width of the decay $a\to 1+2$:
\begin{eqnarray}
  {\rm Im}\;\Pi_b(s_{12}) & = & - \sqrt{s_{12}} \Gamma_b(s_{12}) \ \ , \\
        \Gamma_b(s_{12})  & = & \frac{g_b^2 P_{12}(s_{12})}{8\pi s_{12}}
\end{eqnarray} 
where $g_{b}$ is the coupling constant for the decay $b\to 12$
(for the sake of illustration all particles are assumed to be spinless
in this section), 
and $P_{12}(s_{12})$ is the three momentum of particles 1 and 2 in their CMS:
\begin{eqnarray}
     P_{12}(s_{12}) & = &
     \frac{1}{2} \sqrt{(s_{12}-(m_1+m_2)^2)(1-(m_1-m_2)^2/s_{12})} \ \ . 
\end{eqnarray}
If the resonance $b$ is narrow, then the following
approximation for the resonance propagator can be used:
\begin{eqnarray}
  \frac{1}{p_b^2 - m_0^2 - \Pi_b(p_b^2)} & = &
  \frac{1}{p_b^2 - m_b^2 + i m_b \Gamma_b}
\end{eqnarray}
where $m_b$ and $\Gamma_b=\Gamma_b(m_b^2)$ are the physical mass 
and the physical width of the resonance. 

   The amplitude corresponding to the resonant rescattering diagram in
Fig.\ref{DecayFSI}b has the form
\begin{eqnarray}
    T_{a\to 123}^{res} & = &
    - (i + \beta(s_{12}))
                  \frac{g_a g_b P_{12}(s_{12})}{8\pi \sqrt{s_{12}}}
                  \frac{g_b}{s_{12} - m_b^2 + i m_b \Gamma_b} =  \nonumber \\
    & = &  
    - (i + \beta(s_{12})) \frac{\sqrt{s_{12}}\Gamma_b(s_{12})}
                       {s_{12} - m_b^2 + i m_b \Gamma_b}
\label{Tres}
\end{eqnarray}  
where $g_a$ is the coupling constant for the decay $a\to 123$,
and $\beta(s_{12})$ is the ratio of the real to imaginary part of the loop in
the diagram Fig.\ref{DecayFSI}b.
The imaginary part of the loop, which corresponds to the particles 1 and 2 on
the mass shell (Fig.\ref{DecayFSI}c), is determined by the coupling
constant and the phase space for the decay $b\to 1+2$.
The real part of the loop is divergent and must be regularized,
e.g. by subtraction or by introducing form factors in the vertices:
\begin{eqnarray}
    g_a & \to & g_a F_a(s_{12}) \ \ , \label{FFa} \\
    g_b & \to & g_b F_b(s_{12}) \ \ . \label{FFb}
\end{eqnarray}
The loop in the diagram of Fig.\ref{DecayFSI}b can be
calculated using the dispersion relation
\begin{eqnarray}
   -(i + \beta(s_{12})) \frac{P_{12}(s_{12})}{\sqrt{s_{12}}} & = &
    \frac{1}{\pi}
    \int_{(m_1+m_2)^2}^{\infty}
    \frac{F_a(s) F_b(s) P_{12}(s) s^{-1/2} ds}{s - s_{12} + i\epsilon}
    \ \ . 
\label{DR}
\end{eqnarray}
Instead of specifying the subtraction point or the form factors one can
consider $\beta$ as a model parameter.
Neglecting the real part $(\beta=0)$ corresponds to the unitarity
approximation for $T_{a\to 123}^{UA}$ for the resonant FSI 
amplitude (Fig.\ref{DecayFSI}c).
 
The total amplitude $a\to 123$ has the form 
\begin{eqnarray}
    T_{a\to 123} & = & g_a + T_{a\to 123}^{res} = g_a A(s_{12}) 
\label{Ta123}
\end{eqnarray}
where $A(s_{12})$ is the enhancement factor resulting from
the resonant final state interaction: 
\begin{eqnarray}
    A(s_{12}) & = &
        \left( 1 -  
        \frac{(i + \beta) m_b \Gamma_b}{s_{12} - m_b^2 + i m_b \Gamma_b}
        \right) =
        \frac{s_{12} - m_b^2 - 2m_b\Delta}{s_{12} - m_b^2 + i m_b \Gamma_b}
        \ \ . 
\label{As}
\end{eqnarray}
According to Eqs.(\ref{Ta123},\ref{As}) the total amplitude has a pole at
$s_{12}\approx(m_b-i\Gamma_b/2)^2$ and a {\it zero} at  
$s_{12}\approx(m_b+\Delta)^2$ where
\begin{eqnarray}
      \Delta & = & \beta \Gamma_b/2
\label{Delta}
\end{eqnarray}
The zero of the total amplitude results from the interference between
the resonant term and the nonresonant background,
and it is essential for providing the correct
asymptotic behaviour of the enhancement factor at large $s$:
$A(s)\stackrel{s\to\pm\infty}{\longrightarrow}1$
(see \cite{LMZ97} and references therein). 

The differential decay rate has the form
\begin{eqnarray}
    d\Gamma_{a\to 123} & = &
    (2\pi)^4 \frac{g_a^2}{2m_a} |A((p_1+p_2)^2)|^2 
    d\Phi_3(p_a,p_1,p_2,p_3)
\label{dG123}
\end{eqnarray}
where $d\Phi_3(p_a,p_1,p_2,p_3)$ is the differential 3-body phase space,
see Appendix~A.   
The distribution in the invariant mass of the pair (1+2) is given by
\begin{eqnarray}
    \frac{d\Gamma_{a\to 123}}{d s_{12}} & = &
    (2\pi)^7 \frac{g_a^2}{2m_a} |A(s_{12})|^2
    \Phi_2(m_a,\sqrt{s_{12}},m_3) \Phi_2(\sqrt{s_{12}},m_1,m_2)
\label{dG123ds}
\end{eqnarray}
Here and below $\Phi_n(m,m_1,\ldots,m_n)$ denotes the total $n$-body
phase space, see Appendix~A.   

 The resonance approximation for the production of the particles 1 and 2
corresponds to the case when only the resonant term $T_{a\to 123}^{res}$ in
Eq.(\ref{Ta123}) is taken into account:   
\begin{eqnarray}
    d\Gamma_{a\to b3\to 123} & = &
    \Gamma_{a\to b3} \;
    \frac{m_b \Gamma_b}{|s_{12} - m_b^2 + i m_b \Gamma_b|^2} \;
    \frac{d s_{12}}{\pi}  \label{dGres} \\
    \Gamma_{a\to b3} & = & \int d\Gamma_{a\to b3} = \nonumber \\
    & = &
    (2\pi)^7 \frac{g_a^2}{2m_a} (\pi m_b \Gamma_b) (1+\beta^2)   
    \Phi_2(m_a,m_b,m_3) \Phi_2(m_b,m_1,m_2) 
\label{Gb3}
\end{eqnarray}
where $\beta = \beta(m_b^2)$.  
The ratio of the resonant production rate $\Gamma_{a\to b3}$ to  
the total rate of the decay $a\to 123$ in the plane wave approximation
(without FSI) 
\begin{eqnarray}
    \Gamma_{a\to 123}^{0} & = &
    (2\pi)^4 \frac{g_a^2}{2m_a} \Phi_3(m_a,m_1,m_2,m_3)
\label{G123PWA}
\end{eqnarray}
is given by the formula
\begin{eqnarray}
    \frac{\Gamma_{a\to b3}} {\Gamma_{a\to 123}^{0}} & = &
    (\pi m_b \Gamma_b) (1+\beta^2)   
         \frac{(2\pi)^3\Phi_2(m_a,m_b,m_3)\Phi_2(m_b,m_1,m_2)}
         {\Phi_3(m_a,m_1,m_2,m_3)} = \label{ResDWA3} \\
    & = & 
    (\pi m_b \Gamma_b) (1+\beta^2)
         \frac{\Phi_2(m_a,m_b,m_3)\Phi_2(m_b,m_1,m_2)}
         {{\displaystyle \int} \Phi_2(m_a,m_{12},m_3)
                               \Phi_2(m_{12},m_1,m_2) dm_{12}^2} 
    \ \ . \nonumber
\end{eqnarray}
Note that the off-shell contribution leads to the enhancement factor
$(1+\beta^2)$.  
The resonance approximation (\ref{dGres}) can be used only for $|\beta|\gg
1$, otherwise one cannot neglect the interference of the resonant term with
the background. For example, if $\beta=0$, then the peak structure
arising from the resonance pole is completely suppressed by the zero in the
amplitude (\ref{Ta123},\ref{As}) at $s_{12}=m_b^2$, so that the production
cross section features a dip instead of a peak in the resonance region.  

   The generalization to the four-particle decay $a\to 1234$ with 
final state interaction in the system $(1+2)$ is straightforward.
In particular, the ratio of the resonant production
rate $\Gamma_{a\to b34}$ to the total rate of the decay $a\to 1234$
in the plane wave approximation is given by the formula
\begin{eqnarray}
    \frac{\Gamma_{a\to b34}} {\Gamma_{a\to 1234}^{0}} & = &
         (\pi m_b \Gamma_b) (1+\beta^2) 
         \frac{(2\pi)^3\Phi_3(m_a,m_b,m_3,m_4)\Phi_2(m_b,m_1,m_2)}
         {\Phi_4(m_a,m_1,m_2,m_3,m_4)} = \label{ResDWA4} \\
    & = & 
    (\pi m_b \Gamma_b) (1+\beta^2) 
    \frac{\Phi_3(m_a,m_b,m_3,m_4)\Phi_2(m_b,m_1,m_2)}
    {{\displaystyle \int} \Phi_3(m_a,m_{12},m_3,m_4)
                          \Phi_2(m_{12},m_1,m_2) dm_{12}^2} 
    \ \ . \nonumber
\end{eqnarray}
The differential decay rate has the form
\begin{eqnarray}
    d\Gamma_{a\to 1234} & = &
    (2\pi)^4 \frac{g_a^2}{2m_a} |A((p_1+p_2)^2)|^2 
    d\Phi_4(p_a,p_1,p_2,p_3,p_4) \ \ . 
\label{dG1234}
\end{eqnarray}
Equations (\ref{ResDWA3},\ref{ResDWA4}) have a very simple interpretation:
the rate of resonant production is proportional to the 
fraction of the total phase space which overlaps with the resonant region,
with the extra factor $(1+\beta^2)$ accounting for the off-mass-shell effects.  

  Applying Eq.(\ref{ResDWA3},\ref{ResDWA4}) to the case of $p\bar{p}$
annihilation into $K\bar{K}\rho$ and $K\bar{K}\pi\pi$ 
we get the following estimates for the $\phi$ production: 
\begin{eqnarray}
    \frac{\Gamma_{p\bar{p}\to \phi\rho}}
         {\Gamma_{p\bar{p}\to K\bar{K}\rho}}
    & \sim &
    (\pi m_{\phi} \Gamma_{\phi}) (1+\beta^2) 
    \frac{(2\pi)^3\Phi_2(2m_p,m_{\phi},m_{\rho})\Phi_2(m_{\phi},m_K,m_K)}
    {\Phi_3(2m_p,m_K,m_K,m_{\rho})} = \nonumber \\
    & = & 0.053 (1+\beta^2)  \label{Estphirho}  
    \\
    \frac{\Gamma_{p\bar{p}\to \phi\pi^+\pi^-}}
         {\Gamma_{p\bar{p}\to K\bar{K}\pi^+\pi^-}}
    & \sim &
    (\pi m_{\phi} \Gamma_{\phi}) (1+\beta^2) 
   \frac{(2\pi)^3\Phi_3(2m_p,m_{\phi},m_{\pi},m_{\pi})\Phi_2(m_{\phi},m_K,m_K)}
    {\Phi_4(2m_p,m_K,m_K,m_{\rho})} \nonumber \\
    & = & 0.014 (1+\beta^2)  \label{Estphipipi}  
\end{eqnarray}
Equation (\ref{Estphipipi}) at $\beta=0$ agrees with the estimate given in
\cite{LL}.

\begin{figure}
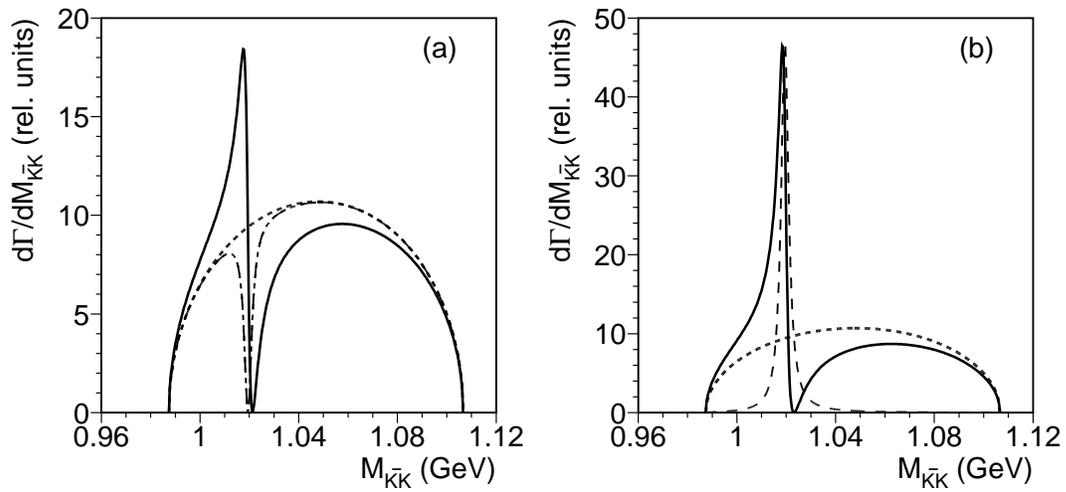

\begin{center}
\mbox{\epsfxsize=70mm \epsffile{OZImKKrhoA.epsf}}
\mbox{\epsfxsize=70mm \epsffile{OZImKKrhoB.epsf}}
\end{center}
\caption{\label{WWM}
The distribution of the invariant mass of the $K\bar{K}$ system for
the reaction $p\bar{p}\to K\bar{K}\rho$ at rest (spinless approximation).  
The full rescattering calculation is given by the solid line:
(a) $\beta=1$, (b) $\beta=2$, 
the phase space distribution by the dotted line. 
The unitarity approximation ($\beta=0$) is shown in (a) by
the dashed-dotted line, the resonance term in (b) by the dashed line.  
All the distributions are nomalized relatively to 
the total rate calculated in the plane wave approximation.}
\end{figure}

To demonstrate the effects of resonant FSI and to explore the difference
between the full loop calculation and the unitarity approximation we use 
the coupled channel model described in Appendix~B. In this
model the loop integrals are regularized by a form factor in the vertex
$b\to 12$. Using a monopole form factor with cutoff parameter $\lambda$
gives a ratio of the real to imaginary part of the loop
$\beta=\lambda/P_{12}$.  As an example we choose the masses and resonance
parameters corresponding to the process $p\bar{p}\to K\bar{K}\rho$ and
calculate the $\phi\rho$ final state (see below for a proper treatment of the
spin factors). The results are shown in Fig.\ref{WWM}.

\section{The $\phi$ meson production in the reaction
            $\bar{p}p\to K\bar{K}\pi^+\pi^-$ at rest}
\label{RRMphipipi}

  In this section we consider $\phi$ meson production in $p\bar{p}$
annihilation at rest in the $K\bar{K}\pi\pi$ channel. The $K\bar{K}$ system
must be in a $P$-wave with $C$-parity $C_{K\bar{K}}=-1$ to form the
$\phi$ meson.  The $\pi\pi$ system can be either in the $C$-odd or the
$C$-even state, and, if we restrict ourselves to $S$-wave annihilation,
the following transitions are possible:
\begin{eqnarray}
    J^{PC}=1^{--}: \ \  p\bar{p}(^3S_1)  & \to & 
    (K\bar{K})_{L=1,P=-1,C=-1} \ (\pi\pi)_{L=even,P=+1,C=+1}  \label{ppbar3S1} 
    \\
    J^{PC}=0^{-+}: \ \  p\bar{p}(^1S_0)  & \to &
    (K\bar{K})_{L=1,P=-1,C=-1} \ (\pi\pi)_{L=odd,P=-1,C=-1} \label{ppbar1S0} 
\end{eqnarray}
The $p\bar{p}(^{2S+1}L_J)K\bar{K}\pi\pi$ vertices with minimum number of
derivatives have the form:
\begin{eqnarray}
    T_{p\bar{p}(1^{--})\to K\bar{K}\pi\pi} & = & 
    g_1 \epsilon^{p\bar{p}}_{\mu} (p_1 - p_2)^{\mu} F_{1}        \label{gS1}
    \\
    T_{p\bar{p}(0^{-+})\to K\bar{K}\pi\pi} & = & 
    g_0 \varepsilon_{\alpha\beta\gamma\delta}
      p_1^{\alpha} p_2^{\beta} p_3^{\gamma} p_4^{\delta} F_{0} \label{gS0}
\end{eqnarray}
where $g_0$ and $g_1$ are the corresponding coupling constants,
$\epsilon^{p\bar{p}}_{\mu}$ is the polarization vector of the 
$p\bar{p}(^3S_1)$ state, the $p_i$ are the four-momenta
of the particles in the final state ($p_1$ and $p_2$ correspond to
$K$ and $\bar{K}$, and $p_3,p_4$ to $\pi\pi$, respectively).
The vertex form factors are denoted by $F_{1}$ and $F_{0}$.
In the case (\ref{gS0}), the strong $\pi\pi$ interaction due to the
$\rho$ meson must be taken into account, using e.g. the method described
in Sec.~\ref{RRM}.
The results for the $p\bar{p}$ annihilation into the $\phi\rho$
channel will be published elsewhere.  


In the rest of this paper we focus our attention on the $\phi$ production
from the triplet $S$-wave $p\bar{p}$ state. The recent partial wave analysis
of the $\phi\pi\pi$ channel measured by OBELIX \cite{OBEL96,OBEL96a}
demonstrates that the $S$-wave reaction is dominated by the $^3S_1$ initial
state if the $\pi\pi$ system is produced with invariant mass below the
$\rho$ resonance, therefore our calculations can be directly compared with
these data\footnote{Due to low acceptance of the $\phi\rho$ channel
the yields of the $\phi\pi^+\pi^-$ channel measured by OBELIX are lower
than those measured by other groups (see Table~\ref{TabpipiKK}).}.
  The coupling (\ref{gS1}) is
proportional to the relative momentum of the $K\bar{K}$ pair,
$P_{K\bar{K}}=\sqrt{(p_K-p_{\bar{K}})^2}/2$, therefore the form factor in
the annihilation vertex is important for damping the transition strength
at large $P_{K\bar{K}}$. We use the following parametrizations: 
\begin{eqnarray}
    F_{1}(P_{K\bar{K}}) & = &
    \left\{
    \begin{array}{lcl} 
    \frac{\displaystyle \Lambda^2}{\displaystyle
                                  \Lambda^2+P_{K\bar{K}}^2} & ~~~~~~~ &
    \mbox{\rm (a) --- monopole} \\ 
    \frac{\displaystyle \Lambda^4}{\displaystyle
                                  (\Lambda^2+P_{K\bar{K}}^2)^2} &  & 
    \mbox{\rm (b) --- dipole} \\ 
    \end{array}
    \right.
\label{AFF}
\end{eqnarray}
where $\Lambda$ is a cut-off parameter which will be determined later. 
We neglect the final $\pi\pi$ interaction in the $S$-wave, since the energy
dependence of the FSI effects below the $f_0(980)$ resonance is known to
be rather smooth (see \cite{LMZ97} and references therein).
For the sake of simplicity we also neglect the FSI effects in the
$K\pi$ systems; this assumption, however, should be removed in further
studies, because the $K^*$ production is rather strong in the $K\bar{K}\pi\pi$
channel \cite{ArREV,Ba66}.

The calculation of the resonant rescattering amplitude is similar
to Sec.\ref{RRM}.  
The result for the differential decay rate is 
\begin{eqnarray}
    \frac{d\Gamma_{K\bar{K}\pi^+\pi^-}}
         {d m_{K\bar{K}}^2}  & = &
    (2\pi)^7 \frac{g_1^2}{2 m_{p\bar{p}}} |A_1(m_{K\bar{K}}^2)|^2
    W(m_{p\bar{p}},m_{K\bar{K}})  \ \ ,     \label{dGKKbar}
    \\
    W(m_{p\bar{p}},m_{K\bar{K}}) & = & 
    \Phi_3(m_{p\bar{p}},m_{K\bar{K}},m_{\pi},m_{\pi})
    \Phi_2(m_{K\bar{K}},m_K,m_{\bar{K}})
    \label{W}
\end{eqnarray}
where $m_{p\bar{p}}=2m_p$,
$m_{K\bar{K}}=\sqrt{s_{K\bar{K}}}=2\sqrt{m_K^2+P_{K\bar{K}}^2}$
is the invariant mass of the $K\bar{K}$ system,
and the enhancement factor $A_1$ has the form 
\begin{eqnarray}
     A_1(s_{K\bar{K}}) & = & f_1(s_{K\bar{K}}) 
        - f_1(m_{\phi}^2) 
        \frac{(i + \beta(s_{K\bar{K}})) m_{\phi} \Gamma_{\phi}}
             {s_{K\bar{K}} - m_{\phi}^2 + i m_{\phi} \Gamma_{\phi}}
        \ \ , \label{A1s}
     \\ 
     f_1(s_{K\bar{K}})  & = & 2 P_{K\bar{K}} F_1(P_{K\bar{K}}) \ \ .
\label{f1}  
\end{eqnarray}
Keeping the resonance term only gives the production rate
\begin{eqnarray}
    \Gamma_{\phi\pi^+\pi^-} & = &
    (2\pi)^7 \frac{g_1^2}{2\sqrt{s}} (\pi m_b \Gamma_b)
    (1+\beta(s_{K\bar{K}})^2)
    f_1(m_{\phi}^2)
    W(m_{p\bar{p}},m_{K\bar{K}}) 
\end{eqnarray}  
which can be compared with the total decay rate to the $K\bar{K}\pi^+\pi^-$
channel in the plane wave approximation:
\begin{eqnarray}
 \frac{\Gamma_{\phi\pi^+\pi^-}}{\Gamma_{K\bar{K}\pi^+\pi^-}^{0}}
    & = &
 (\pi m_b \Gamma_b) (1+\beta^2)
 \frac{f_1(m_{\phi}^2) W(m_{p\bar{p}},m_{\phi})} 
 {{\displaystyle \int} f_1(m_{K\bar{K}}^2)
                       W(m_{p\bar{p}},m_{K\bar{K}}) dm_{K\bar{K}}^2} 
 \ \ . 
\label{f0PWA}
\end{eqnarray}

For an explicit evaluation we proceed by the following steps. First, the
form factor (\ref{AFF}) is parametrized by fitting the experimental 
invariant mass distribution $d\sigma/dm_{K\bar{K}}$ in the mass range
outside the $\phi$ resonance for the reaction $p\bar{p}\to
(K\bar{K})_{L=1}\pi\pi$ at rest. We use the data  on the reaction
$p\bar{p}\to K_{S}K_{L}\pi^+\pi^-$ \cite{Ba66} which selects the final states
$K^0\bar{K}^0$ with $C_{K\bar{K}}=-1$ and $L_{K\bar{K}}=1$. Since the
partial wave analysis for the final state $K_{S}K_{L}\pi\pi$ is not
available, we must rely on some assumptions about the final state
decomposition.
One of them is the dominance of the $S$-wave annihilation,
which allows two possibilities:
$p\bar{p}(^3S_1,1^{--})\to(K^0\bar{K}^0)_{C=-1}(\pi\pi)_{C=+1}$ and    
$p\bar{p}(^1S_0,0^{-+})\to(K^0\bar{K}^0)_{C=-1}(\pi\pi)_{C=-1}$.     
Second, we assume that the contribution from the singlet spin state does not
bring a significant distortion of the shape of $d\sigma/dm_{K\bar{K}}$
outside the $\phi$ meson region. 
The resulting cut-off parameter is 
$\Lambda=0.2\;$GeV for the monopole form factor (\ref{AFF}a) and  
$\Lambda=0.4\;$GeV for the dipole form factor (\ref{AFF}b). 

  After fixing the form factor for the $p\bar{p}$ annihilation vertex,
the only parameter which remains to be determined is the ratio of the
real to imaginary part, $\beta$.  
  For the dipole form factor (\ref{AFF}b) in the annihilation vertex,
the loop in the diagram Fig.\ref{DecayFSI}b is finite, and using
Eq.(\ref{DR}) one gets
\begin{eqnarray}
  \beta(s_{K\bar{K}}) & = &
  \frac{3\Lambda P_{K\bar{K}}^2 + \Lambda^3}{2P_{K\bar{K}}^3}, 
\label{beta}
\end{eqnarray}
leading to $\beta=4.3$ for $\Lambda=0.4\;\mbox{\rm GeV}$.  
  If the monopole form factor (\ref{AFF}a) is used in the annihilation vertex,
then an additional form factor is needed for the
$\phi KK$ vertex.   Introducing a monopole form factor with the cutoff
parameter $\Lambda_{\phi}=\Lambda$ leads again to Eq.(\ref{beta}),
and $\beta=2.0$ at $\Lambda=0.2\;\mbox{\rm GeV}$. 
These estimates show that the real part of the loop is significant even for
rather soft form factors because the imaginary part is suppressed by the
factor $P_{K\bar{K}}^3$ corresponding to the $P$-wave of the
relative motion in the $K\bar{K}$ system.    

\begin{figure}
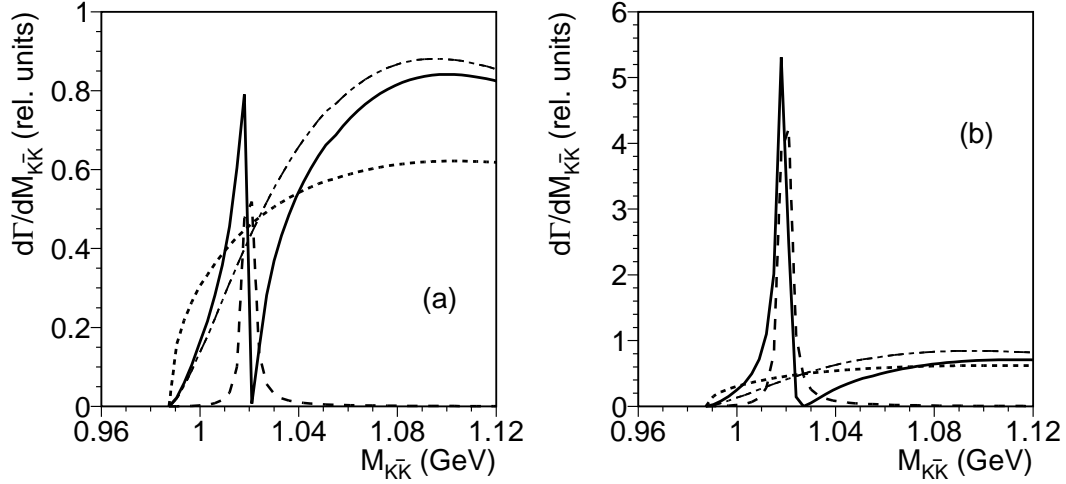

\begin{center}
\mbox{\epsfxsize=70mm \epsffile{OZImKKpipiA.epsf}}
\mbox{\epsfxsize=70mm \epsffile{OZImKKpipiB.epsf}}
\end{center}
\caption{\label{FigKKpipi}
The distribution of the invariant mass $m_{K\bar{K}}$ for the reaction 
$\bar{p}p(^3S_1)\to K\bar{K}\pi^+\pi^-$:
(a) $\beta=1$, (b) $\beta=4$. 
The full calculation is given by the solid line,
the resonance term only by the dashed line,
the plane wave approximation by the dash-dotted line,
the phase space distribution by the dotted line.}
\end{figure}

\begin{figure}
\begin{center} \mbox{\epsfysize=80mm \epsffile{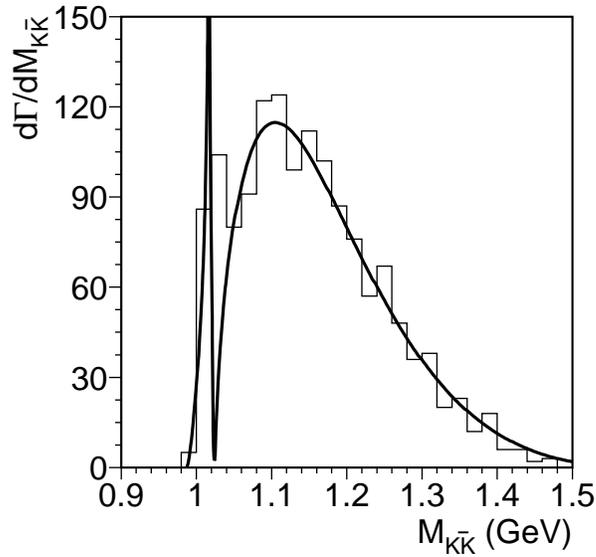}} \end{center}
\caption{\label{FigKKpipiEXP}
The calculated distribution of the invariant mass $m_{K\bar{K}}$
for the reaction $\bar{p}p(^3S_1)\to K\bar{K}\pi^+\pi^-$ ($\beta=2$) 
in comparison with the experimental data \protect\cite{Ba66}.} 
\end{figure}

\begin{table}
\caption{\label{BRpeak}
The calculated ratio of the $\phi$ resonance peak to the total yield 
for the reaction $p\bar{p}(^3S_1)\to (K\bar{K})_{L=1}(\pi\pi)_{L=0}$ at rest
for various values of $\beta$, Eq.(\protect\ref{Tres}). }
\begin{center}
\begin{tabular}{|c|c|c|c|c|}
\hline\hline
   $\beta$            &  1.0    &  2.0    &  3.0    &  4.0  \\
\hline
 dipole form factor   &  0.046  &  0.083  &  0.139  &  0.211 \\
\hline
 monopole form factor &  0.059  &  0.105  &  0.173  &  0.261 \\
\hline\hline
\end{tabular}
\end{center}
\end{table}

The results for the distribution in the invariant mass of the $K\bar{K}$
system are shown in Fig.\ref{FigKKpipi} for various values of $\beta$.
As already discussed in Sec.~\ref{RRM}, the resonant FSI mechanism 
causes a characteristic peak-dip structrure in the invariant mass
distribution around the $\phi$ mass.  
This feature can be used to distinguish the resonant FSI mechanism from
other mechanisms of $\phi$ production, including the intrinsic
strangeness \cite{EGK,EKKS} and the two-meson doorway mechanism
$p\bar{p}\to\rho\omega\to\phi\pi\pi$ \cite{LL}, which give only a peak
in the region of the $\phi$ meson.    
The calculated mass distribution for $\beta=2$ is compared with the
experimental data in Fig.\ref{FigKKpipiEXP}.
A lower limit for $\beta$ can, in principle, be obtained from the experimental
data using Eq.(\ref{Delta}) provided the position of the zero is known. 
This requires the separation of different partial waves in the distribution
of the effective mass of the $K\bar{K}$ system, which is not presently
available. However, from the fact that the resonance peak in the experiment
\cite{OBEL96,OBEL96a} is observed very close to the $\phi$ mass, i.e. its
position is not shifted due to a nearby zero, we can conclude that
$\beta\geq2$.
Thus both theoretical estimates and the data indicate a large value of the
real part of the loop.
The calculated ratio of the $\phi$ resonance peak to the total rate of the
$(K\bar{K})_{L=1}(\pi\pi)_{L=0}$ channel is shown in Table~\ref{BRpeak} (the
peak contribution is defined as the integral from the $K\bar{K}$ threshold
to the zero of the production amplitude, see
Figs.\ref{FigKKpipi},\ref{FigKKpipiEXP}).  

To calculate the absolute yield of the $\phi\pi\pi$ we use the measured
branching ratio $BR(p\bar{p}\to K_SK_L\pi^+\pi^-)=2.4\cdot 10^{-3}$
\cite{Ba66} assuming a fraction of $3/4$ for to the
annihilation from the $^3S_1$ state. Considering all charge channels in
the intermediate state doubles the branching ratio leading to our estimate 
$BR(p\bar{p}\to (K\bar{K})_{L=1}\pi^+\pi^-)=3.6\cdot 10^{-3}$.
Using the results for the relative yield of the $\phi$ meson from
Table~\ref{BRpeak} gives the absolute yield to the $\phi\pi^+\pi^-$ channel
for the $p\bar{p}$ annihilation at rest as  
$BR(p\bar{p}(^3S_1)\to\phi\pi^+\pi^-)\geq 3\cdot 10^{-4}$ for $\beta\geq 2$. 
This result agrees well with the OBELIX result for the $\phi\pi\pi$ yield
in liquid $BR(\phi\pi^+\pi^-)=3.5(4)\cdot 10^{-4}$ \cite{OBEL96} (this value
practically excludes the contribution from the $\phi\rho$ channel).
The other measurements of $BR(\phi\pi^+\pi^-)$ shown in
Table~\ref{TabpipiKK} give slightly higher values due to the $\phi\rho$
contribution. 
  The ASTERIX analysis \cite{AST91} gives 
$BR(p\bar{p}(S)\to\phi\pi^+\pi^-)=4.7(11)\cdot 10^{-4}$ and 
$BR(p\bar{p}(^1S_0)\to\phi\rho)=3.4(10)\cdot 10^{-4}$.
Using these branching ratios and the fact that the singlet $S$-wave
corresponds to about one quarter of the total annihilation in liquid one
gets $BR(p\bar{p}(^3S_1)\to\phi\pi^+\pi^-)=BR(p\bar{p}(S)\to\phi\pi^+\pi^-)
-\frac{1}{4}BR(p\bar{p}(^1S_0)\to\phi\rho)=3.8(13)\cdot 10^{-4}$, 
in agreement with the OBELIX measurement as well as with our calculations.
  Note, that the old bubble chamber measurement \cite{Ba66}
$BR(p\bar{p}(S)\to\phi_{\to K_SK_L}\pi^+\pi^-)=1.8(3)\cdot 10^{-4}$ gives 
$BR(p\bar{p}(S)\to\phi\pi^+\pi^-)=
BR(p\bar{p}(S)\to K_SK_L\pi^+\pi^-)/BR(\phi\to K_SK_L)=5.4(9)\cdot 10^{-4}$
in good agreement with the LEAR results.


   The two-step mechanism considered is one of many possible two-step
processes. Other four-particle intermediate states to be discussed include
the  $K^*\bar{K}\pi\pi$, $K\bar{K}^*\pi\pi$, and $K^*\bar{K}^*\pi\pi$ states
which arguably can lead to Lipkin cancelations \cite{Li84}.
However, because the branching ratio for these channels is much smaller than
for the $K\bar{K}\pi\pi$ channel and the dispersion integral (\ref{DR}) is
saturated mainly  in the low $s$ region due to the relatively soft form
factors, we do not expect strong cancellations between the contributions from
these additional four-particle states.
   We do not include the $\phi-\omega$ mixing term in order to avoid
possible double counting%
\footnote{The $\phi-\omega$ mixing goes via a two-step mechanism
with the $K\bar{K}$, $K^*\bar{K}$, $K\bar{K}^*$, $K^*\bar{K}^*$ intermediate
states.}.

   The contribution of the $\rho\omega$ intermediate state, which is
potentially significant \cite{LL}, remains to be calculated, but there is no
reason to expect that there is a cancelation between the $\rho\omega$ and
$K\bar{K}\pi\pi$ terms (their relative phase is an unknown parameter).  

   Using Eq.(\ref{dGKKbar}) one can estimate the dependence of the relative
$\phi\pi\pi$ yield on the invariant mass of the $p\bar{p}$. For the sake of
simplicity we use the phase space distribution for the $\pi\pi$ pair and the
form factors for the $(K\bar{K})_{L=1}$ system described above.  The results
shown in Table~\ref{BRvsE} demonstrate that the fraction of the total phase
space of the $K\bar{K}\pi\pi$ system, which is favourable for the
resonant $\phi$ production, decreases with increasing total energy.
This is in qualitative agreement with the general trend seen in OZI rule
violation in $p\bar{p}$ annihilation%
\footnote{To obtain the absolute branching ratio for the $\phi\pi\pi$
channel from  the results shown in Table~\ref{BRvsE}, the energy dependence
of the vertex $p\bar{p}\to (K\bar{K})_{L=1}\pi\pi$, which is experimentally
unknown, must be taken into account. Because the $K$ and $\bar{K}$ share
the same $s$-quark line, the $K\bar{K}$ pairs are predominantly produced
with limited relative orbital momentum, and the energy dependence of the
relative $P$-wave $K\bar{K}$ production is not expected to be strong.}. 

\begin{table}
\caption{\label{BRvsE}
The calculated ratio of the $\phi$ resonance peak to the total yield 
$BR(\phi\pi\pi)/BR((K\bar{K})_{L=1}\pi\pi)$ 
for the reaction $p\bar{p}\to (K\bar{K})_{L=1}\pi\pi$ 
as a function of the total energy.} 
\begin{center}
\begin{tabular}{|c|cccc|}
\hline\hline
 $\sqrt{s_{p\bar{p}}}/m_p$  &   2    &  2.5   &  3      &   4     \\
\hline
  dipole form factor, $\beta=2$ & 
                               0.083 &  0.050 &  0.039  &  0.031  \\  
  monopole form factors, $\beta=2$ & 
                               0.105 &  0.058 &  0.041  &  0.027  \\  
\hline\hline
\end{tabular}
\end{center}
\end{table}

\section{Conclusion}
\label{CONCL}

  We have developed a general formalism describing resonant final state
interaction in a multiparticle system. This mechanism is shown to play an
important role in nucleon-antinucleon annihilation into OZI-rule
breaking channels ($\phi\pi\pi$, $\phi\rho$). 
  The resonant rescattering mechanism 
$\bar{p}p\to\pi^+\pi^-K\bar{K}\to\pi^+\pi^-\phi$ is studied in detail. 
Off-mass-shell contributions are found to be very significant for this
process, while the unitarity approximation is small.  
 The interference of the resonant term with the nonresonant background 
which is essential for providing a correct analytical structure of the total
amplitude in the case of elastic resonant rescattering, leads to the
characteristic peak-dip structure of the invariant $K\bar{K}$ mass 
distribution for $J^P_{K\bar{K}}=1^-$. This feature can help to 
distinguish the resonant FSI mechanism from other mechanisms of $\phi$
production when high resolution partial wave analysis of the $K\bar{K}$
production becomes available. 
The $\phi$ peak can get an additional enhancement due to this interference
effect.  
The rate of the $\phi\pi\pi$ production at rest from the $^3S_1$ state
is in good agreement with the experimental data
\cite{AST91,OBEL96,Biz69}.
  Thus no unexplained OZI rule violation is required in this case --- 
the same conclusion was obtained earlier for the $\phi\pi^0$, $\phi\phi$,
$\phi\gamma$ channels as well (see \cite{Lo95,Zou96,Ma96} and references
therein). 
  The $\phi$ production due to resonant FSI decreases with increasing total
energy as the fraction of the total phase space 
favourable for the resonance formation gets smaller. This feature agrees
with the general trend of the OZI rule violation in the nucleon-antinucleon
annihilation which is decreasing as a function of energy. 

  It would be desirable to extend the present approach to the case when
resonant FSI effects are taken into account simultaneously in the different
subsystems mentioned earlier in order to achieve a unified description of
the production of the $\phi$, $\rho$, and $K^*$ resonances.

\section*{Acknowledgments}

We are grateful to M.~Sapozhnikov and B.-S.~Zou for stimulating
discussions.

\appendix
\section{Appendix: Phase Space} \label{APA2}

  The differential $n$-particle phase space
$d\Phi_n(p,p_1,p_2,\ldots,p_n)$ is defined by
\begin{eqnarray}
 d\Phi_n(p,p_1,p_2,\ldots,p_n) & = & (2\pi)^{-3n}
       \delta^4(p -\sum_{i=1}^n p_i) \prod_{i=1}^n
       \frac{d^3\mbox{\bf p}_i}{2E_i}
\end{eqnarray}
where $p$ is the total four-momentum of the particles with
four-momenta $p_i=(E_i,\mbox{\bf p}_i)$. 
  The phase space reduction formula has the form 
\begin{eqnarray}
  d\Phi_n(p,p_1,p_2,\ldots,p_n) & = &
  d\Phi_{n-1}(p,p_x,p_3,\ldots,p_n) d\Phi_2(p_x,p_1,p_2) (2\pi)^3 dp_x^2 
\label{Phi4}
\end{eqnarray}
The total $n$-particle phase space for the decay $a\to 12\ldots n$ is
\begin{eqnarray}
  \Phi_n(m_a,m_1,m_2,\ldots,m_n) & = &
  \int d\Phi_n(p,p_1,p_2,\ldots,p_n) \ \ , \ p^2=m_a^2 \ \ . 
\end{eqnarray}

\section{Appendix: Resonant FSI in a Coupled Channel Model} \label{APA}

In order to describe the resonant final state interaction in a two-particle
system, we introduce a variant of the Weisskopf-Wigner (WW) model with two
channels. Channel 1 is the scattering channel of interest, and channel 2 has
a bound state $|b\rangle$ with bare mass $m_0^2$ (the rest of the
dynamics in the second channel is ignored).  The only interaction in the
model results from the coupling between the channels.
  The $T$-matrix, as a function of the invariant mass squared $s$, is
defined by the Lippmann-Schwinger equation
\begin{eqnarray}
   \left(\matrix{T_{11} & T_{12} \cr T_{21} & T_{22} \cr} \right) 
   & = &
     \left(\matrix{ 0 & V \cr V^+ & 0 \cr}\right) +  
     \left(\matrix{ 0 & V \cr V^+ & 0 \cr}\right)
     \left(\matrix{G^0_1(s) & 0 \cr 0 & G^0_2(s) \cr} \right) 
     \left(\matrix{T_{11} & T_{12} \cr T_{21} & T_{22} \cr} \right) 
\label{T}
\end{eqnarray}
where $V$ is the interaction between channels 1 and 2 and
$G^0_1(s)$ and $G^0_1(s)$ are the free Green functions: 
\begin{eqnarray}
  G^0_{1}(s) & = & \frac{2}{\pi}\;
  \int_{0}^{\infty} \frac{|k \rangle\langle k |}{s/4-m^2-k^2)}
                    k^2 dk  \label{G0k}
  \\ 
  G^0_{2}(s) & = &
  \frac{|b \rangle\langle b| }{s-m_0^2} \ \ . \label{G0b}
\end{eqnarray}
Here $|k\rangle$ denotes the free two-particle state with relative
momentum $k$, both particles have the same mass $m$, and  
$s=4(k^2+m^2)$. 
  We assume the channel coupling to have the following form
(all particles are spinless)  
\begin{equation}
  \langle k | V | b \rangle = g \xi(k) = \frac{g}{k^2+\mu^2}
\end{equation}
where $g$ is the coupling constant and $\mu$ characterizes the range
of interaction.

The solution for the scattering amplitude in channel 1 has the form
\begin{eqnarray}
   f(s) & = &
   - \langle k | T_{11}(s) | k \rangle =
   - \;\frac{g^2 \xi(k)^2 }{s - m_0^2 - \Pi(s)}
\label{f11}
\end{eqnarray}
where the mass operator $\Pi(s)$ is given by 
\begin{eqnarray}
   \Pi(s) & = & g^2 \langle \xi | G_1^0(s) | \xi \rangle =  
   \frac{g^2 }{2\mu(k+i\mu)^2}  \ \ . 
\label{Pis}
\end{eqnarray}
From Eqs.(\ref{f11},\ref{Pis}) the poles of the amplitude can be easily
found. For the sake of illustration we consider the case of weak coupling
(the bare state $b$ is assumed to be above the threshold of the scattering
channel), then the scattering amplitude has a resonance pole at 
$s = (m_b - i\Gamma_b/2)^2 \approx m_0^2 + \Pi(m_0^2)$. 
The resonance shift $\delta m_b = m_b-m_0$ and width $\Gamma_b$
due to the coupling to the open channel have the form
\begin{eqnarray}
    \delta m_b & \approx & \frac{\mbox{\rm Re}\;\Pi(m_0)}{2 m_0} \\    
    \Gamma_b   & \approx & -\frac{\mbox{\rm Im}\;\Pi(m_0)}{m_0} =
                   \frac{2 g^2 k_b^2 \xi^2(k_b)}{m_0}    
\end{eqnarray}
where $k_b=\sqrt{m_b^2/4-m^2}$. 

  Now we consider the decay $a\to 123$ described by a pointlike vertex with
a coupling constant $g_a$. Using standard results from scattering
theory we find that the decay amplitude with the final state interaction
between particles 1 and 2 taken into account is proportional to the
scattering wave function at zero distance $r_{12}$ between
particles 1 and 2. The result for the above described model has the form
\begin{eqnarray}
   F_{a\to 123} & = &
    g_a \langle r_{12}=0 | k_{12}^{(+)} \rangle =  \nonumber \\ 
    & = & g_a \langle r_{12}=0 | k_{12} \rangle +
        \langle r_{12}=0 |G_0(s_{12}) T(s_{12})| k_{12} \rangle \\
    & = & g_a \left( 1 - \frac{g^2 (\mu+ik_{12}) \xi^2(k_{12})}
                     {s_{12} - m_0^2 - \Pi(s_{12})} \right) 
\label{Fa123}
\end{eqnarray}
where the relative momentum $k_{12}$ is related to the the square of the
invariant mass of the subsystem (1+2):
\begin{eqnarray}
   s_{12} & = & 4(m^2+k_{12}^2)  \ \ . 
\end{eqnarray} 

Formula (\ref{Fa123}) can be rewritten explicitly showing the 
pole-zero structure of the amplitude:
\begin{eqnarray}
   F_{a\to 123} & = & g_a
   \frac{s_{12} - m_0^2 - \Delta(s_{12})}{s_{12} - m_0^2 - \Pi(s_{12})}
   \label{FABCCM} \\  
   \Delta(s_{12}) & = & \frac{g^2 }{2\mu} \frac{1}{(k_{12}^2+\mu^2)}
   \ \ . \label{DeltaCCM} 
\end{eqnarray}
Note that the nominator in (\ref{FABCCM}) is a real function for real
$s_{12}$ which has a zero at $s_{12} = m_z^2 \approx m_0^2+\Delta(m_b^2)$.
The distance between the zero and the pole in the weak coupling limit is  
\begin{eqnarray}
    m_z - m_b & = & \frac{\mu \Gamma_b}{2 k_b}  \ \ . 
\end{eqnarray}



\end{document}